\documentclass[aps,prb,preprint,showpacs,superscriptaddress,amsmath,amssymb]{revtex4}

\usepackage{graphicx}
\usepackage{dcolumn}
\usepackage{bm}
\usepackage{color}

\usepackage{amssymb, amsmath}

\begin{document}

\title{Superposition principle for nonlinear Josephson plasma waves in layered superconductors}

\author{T.N.~Rokhmanova}
\affiliation{A.Ya.~Usikov Institute for Radiophysics and Electronics NASU, 61085 Kharkov, Ukraine}
\email{Rokhmanova@i.ua}

\author{S.S.~Apostolov}
\affiliation{A.Ya.~Usikov Institute for Radiophysics and Electronics NASU, 61085 Kharkov, Ukraine}
\affiliation{V.N.~Karazin Kharkov National University, 61077 Kharkov, Ukraine}
\email{stapos@ukr.net}

\author{Z.A.~Maizelis}
\affiliation{A.Ya.~Usikov Institute for Radiophysics and Electronics NASU, 61085 Kharkov, Ukraine}
\affiliation{V.N.~Karazin Kharkov National University, 61077 Kharkov, Ukraine}
\email{mjkp@ukr.net}

\author{V.A.~Yampol'skii}
\affiliation{A.Ya.~Usikov Institute for Radiophysics and Electronics NASU, 61085 Kharkov, Ukraine}
\affiliation{V.N.~Karazin Kharkov National University, 61077 Kharkov, Ukraine}
\affiliation{CEMS, RIKEN, Saitama 351-0198, Japan}
\email{yam@ire.kharkov.ua}

\author{Franco Nori}
\affiliation{CEMS, RIKEN, Saitama 351-0198, Japan}
\affiliation{University of Michigan, Ann Arbor, Michigan 48109, USA}
\email{fnori@riken.jp}

\begin{abstract}
We show that a specific superposition principle is valid for \emph{nonlinear} Josephson plasma waves in layered superconductors. We study theoretically the reflection and transmission of terahertz waves through a finite-size superconducting slab placed inside a rectangular waveguide with ideal-metal walls. We assume that the superconducting layers are parallel to the waveguide axis. We show that there exist two specific mutually-orthogonal polarizations for waves which, in spite of the nonlinearity, reflect and transmit through the superconductor \emph{independently}. The wave of the first polarization causes a strong shielding current along the crystallographic \textbf{ab}-plane of the superconductor. Therefore, this wave reflects nearly completely from the superconductor and excites only an evanescent mode inside it. The wave of the other polarization does not contain the electric field component parallel to both the sample surface and the crystallographic \textbf{ab}-plane, and excites much weaker shielding currents. Therefore, it partially reflects and partially transmits through the sample. Moreover, this wave excites the nonlinear mode in the layered superconductor, and the transmission coefficient of the superconductor depends on the amplitude of the incident wave of this polarization. However, this transmission coefficient is independent of the amplitude of the wave with first polarization. On the basis of the discussed superposition principle, we suggest a new method for solving nonlinear problems of waves interaction in layered superconductors. Namely, it is reasonably to represent incident, reflected, and transmitted waves of any polarizations as superpositions of the modes with the two specific polarizations considered here, and then solve the problem separately for these modes. We apply this method to the case of nonlinear interaction and mutual transformation of the transverse electric and transverse magnetic modes in layered superconductors.
\end{abstract}

\pacs{74.78.Fk, 74.50.+r, 74.72.-h}




\maketitle

\section{Introduction}

Experimental studies (see, e.g., Refs.~\onlinecite{Kl-Mu,Kl-Mu2}) proved that layered superconductors can be treated as periodic structures where thin superconducting layers (with thicknesses $s$ of about 0.2~{nm}) are coupled through thicker dielectric layers (with thicknesses $d$ of about 1.5~{nm} and a dielectric constant $\varepsilon \sim 15$) via the \emph{intrinsic Josephson effect}. Strongly anisotropic high-temperature superconductor crystals $\rm Bi_2Sr_2CaCu_2O_{8+\delta}$ or artificial compounds $\rm Nb /Al$~-~${\rm  Al O}_x\rm / Nb$ are examples of such materials. They are of great interest from both technological and fundamental-science view points. For fundamental science, the interest in layered superconductors is related to the specific type of plasma formed there due to its layered structure, the so-called Josephson plasma. Because of the essential difference in the nature of the currents along and across layers, this plasma is strongly anisotropic. Indeed, the strong currents along the layers are of the same nature as in bulk superconductors, while the relatively small currents across the layers are related to the Josephson effect. The strong current anisotropy provides the possibility for propagation of specific Josephson plasma electromagnetic waves (JPWs) in layered superconductors (see, e.g., Refs.~\onlinecite{Thz-rev,rev2} and references therein). JPWs belong to the terahertz frequency range, which is important for various applications, but not easily reachable with modern devices.

The electrodynamics of layered superconductors is described by a set of coupled sine-Gordon equations~\cite{sine-gord,SG2,SG3,SG4,SG5,Thz-rev} for the gauge-invariant interlayer phase difference $\varphi$ of the order parameter. These equations are nonlinear due to the nonlinear relation $J\propto\sin\varphi$ between the Josephson interlayer current $J$ and  $\varphi$.
Here we will consider the case of weak nonlinearity, $|\varphi| \ll 1$, when $\sin\varphi$ can be expanded as $\sin\varphi \approx \varphi-\varphi^3/6$. As was shown in Refs.~\onlinecite{nl1,nl2,nl3,nl4}, even in this case the non-trivial nonlinear phenomena accompanying the propagation of JPWs can be observed, e.g., slowing down of light~\cite{nl1}, self-focusing of terahertz pulses~\cite{nl1,nl2}, excitation of nonlinear waveguide modes~\cite{nl3}, and self-induced transparency of the layered superconductors~\cite{nl4}. The noticeable change in the transparency of the cuprate superconductor, when increasing the wave amplitude, was recently observed in the important experiment, Ref.~\onlinecite{dienst13}, where the excitation of Josephson plasma solitons led to an effective decrease of the Josephson resonance frequency.

\begin{figure}[h]
\begin{center}
\includegraphics [width=16cm]{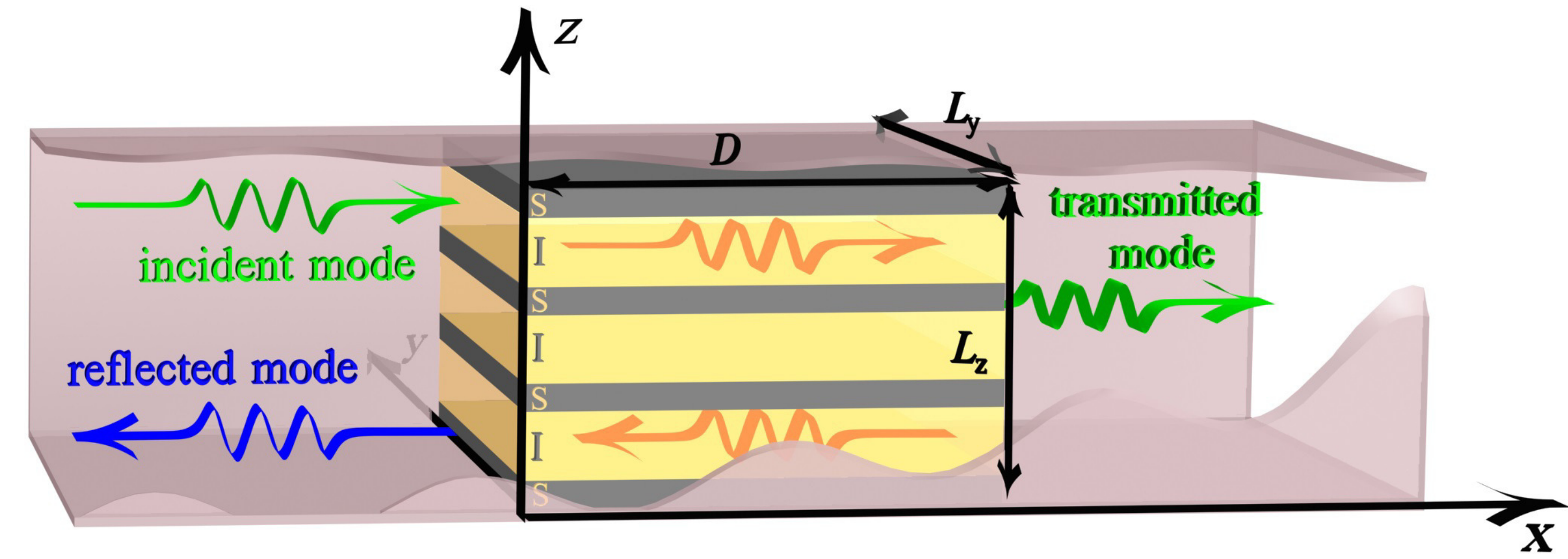}
\caption{(Color online) Schematic geometry for waves propagating in a waveguide along the superconducting layers. Note that here S and I stand for superconducting and insulator layers, respectively. The light pink translucent layer (cut-off to show the sample inside) represents the walls of the waveguide.} \label{wavegAB}
\end{center}
\end{figure}

In this paper, we study theoretically the nonlinear interaction of the electromagnetic waves with different polarizations in a slab of layered superconductor placed into a waveguide with ideal metal walls. The waveguide axis is assumed to be parallel to the superconducting layers (see Fig.~\ref{wavegAB}). Due to the strong current anisotropy of the layered superconductor-vacuum interface (the $yz$-plane), the transformation of the polarizations can be observed for the reflected and transmitted waves. We calculate the dependence of the transformation coefficients on the amplitude of the incident wave for the cases when the transverse magnetic (TM) mode transforms to the transverse electric (TE) mode and vice versa. The main result of our paper consists in  revealing a specific superposition principle which is valid even in the \emph{nonlinear} case. We introduce two special mutually-orthogonal wave polarizations matched with the $y$-axis, which is perpendicular both to the waveguide axis ($x$-axis) and the crystallographic \textbf{c}-axis of the layered superconductor ($z$-axis). The magnetic field in the wave of the first polarization (we call it as H$_\perp$ polarization) and the electric field in the wave of the second polarization (we call it as E$_\perp$ polarization) are perpendicular to the $y$-axis. We show that, in the main order with respect to the anisotropy parameter, the waves of these polarizations interact with the layered superconductor \emph{independently}. The incident wave of the H$_\perp$ polarization has the electric field component parallel both to the vacuum-superconductor interface and to the crystallographic \textbf{ab}-plane. This wave  excites strong shielding currents along the layers and, therefore, it penetrates into the sample over short length in the form of an evanescent wave and reflects almost completely from the superconductor. At the same time, the wave of the E$_\perp$ polarization does not contain the electric field component parallel both to the sample surface and to the crystallographic \textbf{ab}-plane. Therefore, the shielding currents flow along the \textbf{c}-axis only and they are relatively small in this case. So, the wave of the E$_\perp$ polarization can propagate in a layered superconductor and penetrates over long distances. This wave partially reflects and partially transmits through the sample. We show that, in spite of the nonlinearity, the H$_\perp$ and E$_\perp$ waves do not practically interact. Therefore, to study the reflection and transmission of the TE and TM incident waves (or the waves of any other polarization), we can perform the following steps:
\begin{enumerate}
\item Express the incident wave as the \textit{superposition} of two modes of H$_\perp$ and E$_\perp$ polarizations.
\item Study the reflection and transmission of these modes \textit{separately}.
\item Represent the reflected and transmitted fields of the H$_\perp$ and E$_\perp$ modes as superpositions of the TE and TM modes.
\end{enumerate}

This paper is organized as follows. In the next section, we derive the electromagnetic fields in the vacuum and in the layered superconductor and justify the superposition principle for the waves with H$_\perp$ and E$_\perp$ polarizations. In the third section, we study the nonlinear reflection and transmission of the E$_\perp$ waves through the slabs of layered superconductors. In the fourth section, we apply the revealed superposition principle for the cases of the TE and TM waves incidence. In the final section, we summarize the results obtained in the paper.

\section{Electromagnetic field distribution}

\subsection{Geometry of the problem}

Consider a waveguide of lateral sizes $L_y$ and $L_z$ with a sample of layered superconductor of length $D$ inside it (see Fig.~\ref{wavegAB}). The coordinate system is chosen in such a way that the crystallographic $\mathbf{ab}$-plane of the layered superconductor coincides with the $xy$-plane, and the $\mathbf{c}$-axis is along the $z$-axis. An electromagnetic mode of frequency $\omega$ propagates in the waveguide along the $x$-axis, which is parallel to the superconducting layers. The incident wave partly reflects from the layered superconductor and partly transmits through it, as is shown schematically in Fig.~\ref{wavegAB}.

The electric ${\vec E(\vec{r},t)}$ and magnetic ${\vec H(\vec{r},t)}$ fields in the waveguide can be expressed via the vector potential ${\vec A(\vec{r},t)}$ by the usual equations,
\begin{equation}\label{HE}
{\vec H}(\vec{r},t) = {\rm rot} \left[{\vec A}(\vec{r},t)\right], \quad {\vec E}(\vec{r},t)= -
\frac{1}{c}\frac{\partial {\vec A}(\vec{r},t)}{\partial t}.
\end{equation}
The scalar potential is assumed to be equal to zero.

Using the boundary conditions (the tangential components of the electric field are zero on the waveguide walls), we present the components of the vector potential in the following form:
\begin{gather}
A_x(\vec{r},t)=\mathcal{A}_x(x,t) \sin(k_y y)\sin(k_z z),
\notag\\
A_y(\vec{r},t)=\mathcal{A}_y(x,t) \cos(k_y y)\sin(k_z z),
\label{A_mult-s}\\
A_z(\vec{r},t)=\mathcal{A}_z(x,t) \sin(k_y y)\cos(k_z z),
\notag
\end{gather}
where $k_y=\pi n_y/L_y$, $k_z=\pi n_z/L_z$; $n_y$ and $n_z$ are positive integer numbers that define the propagating mode in the waveguide.

\subsection{Electromagnetic field in the vacuum}

The electromagnetic field in the vacuum can be presented as a superposition of waves of the H$_\perp$ and E$_\perp$ polarizations. In the H$_\perp$-polarized wave, the magnetic field is perpendicular to the $y$-axis,
\begin{equation}\label{pol_I}
{\vec E}^{(1)} = \{E_x^{(1)}, E_y^{(1)}, E_z^{(1)}\}, \quad {\vec H}^{(1)} = \{H_x^{(1)}, 0, H_z^{(1)}\},
\end{equation}
whereas the electric field is orthogonal to the $y$-axis in the wave of the E$_\perp$ polarization,
\begin{equation}\label{pol_II}
{\vec E}^{(2)} = \{E_x^{(2)}, 0, E_z^{(2)}\}, \quad {\vec H}^{(2)} = \{H_x^{(2)}, H_y^{(2)}, H_z^{(2)}\}.
\end{equation}
Hereafter, the superscripts (1) and (2) denote the H$_\perp$ and E$_\perp$ polarizations, respectively.

The incident and reflected modes of the H$_\perp$ and E$_\perp$ polarizations propagate in the vacuum region $x<0$. The Maxwell equations give the following expressions for the amplitudes
$\vec {\mathcal{A}}_{\rm inc}(x,t)$ of the vector potential of the incident wave:
\begin{eqnarray}
\mathcal{A}_{x\,{\rm inc}}(x,t)&=&-H_{\rm inc}^{(1)}\dfrac{k_xk_y}{k^3} \sin(k_xx-\omega t+\phi_{\rm inc}^{(1)})
\notag\\
&&-H_{\rm inc}^{(2)}\dfrac{k_z}{k^2} \sin(k_xx-\omega t+\phi_{\rm inc}^{(2)}),
\notag\\
\mathcal{A}_{y\,{\rm inc}}(x,t)&=&-H_{\rm inc}^{(1)}\dfrac{k^2-k_y^2}{k^3} \cos(k_xx-\omega t+\phi_{\rm inc}^{(1)}),
\label{A_inc}
\\
\mathcal{A}_{z\,{\rm inc}}(x,t)&=&H_{\rm inc}^{(1)}\dfrac{k_yk_z}{k^3} \cos(k_xx-\omega t+\phi_{\rm inc}^{(1)})
\notag\\
&&-H_{\rm inc}^{(2)}\dfrac{k_x}{k^2} \cos(k_xx-\omega t+\phi_{\rm inc}^{(2)}),
\notag
\end{eqnarray}
where $H_{\rm inc}^{(1)}$, $\phi_{\rm inc}^{(1)}$, $H_{\rm inc}^{(2)}$, and $\phi_{\rm inc}^{(2)}$ are the amplitudes and phases of the magnetic field for the incident waves of the H$_\perp$ and E$_\perp$ polarizations.

Similar expressions can be written for the vector potential of the reflected waves,
\begin{eqnarray}
\mathcal{A}_{x\,{\rm ref}}(x,t)&=&-H_{\rm ref}^{(1)}\dfrac{k_xk_y}{k^3} \sin(k_xx+\omega t-\phi_{\rm ref}^{(1)})
\notag\\
&&+H_{\rm ref}^{(2)}\dfrac{k_z}{k^2} \sin(k_xx+\omega t-\phi_{\rm ref}^{(2)}),
\notag\\
\mathcal{A}_{y\,{\rm ref}}(x,t)&=&-H_{\rm ref}^{(1)}\dfrac{k^2-k_y^2}{k^3} \cos(k_xx+\omega t-\phi_{\rm ref}^{(1)}),
\label{A_ref}
\\
\mathcal{A}_{z\,{\rm ref}}(x,t)&=&H_{\rm ref}^{(1)}\dfrac{k_yk_z}{k^3} \cos(k_xx+\omega t-\phi_{\rm ref}^{(1)})
\notag\\
&&+H_{\rm ref}^{(2)}\dfrac{k_x}{k^2} \cos(k_xx+\omega t-\phi_{\rm ref}^{(2)}),
\notag
\end{eqnarray}
where $H_{\rm ref}^{(1)}$, $\phi_{\rm ref}^{(1)}$, $H_{\rm ref}^{(2)}$, and $\phi_{\rm ref}^{(2)}$ are the amplitudes and phases of the magnetic field for the reflected waves of the H$_\perp$ and E$_\perp$ polarizations.

In the second vacuum region, at $x>D$, the transmitted waves of the H$_\perp$ and E$_\perp$ polarizations propagate. Their vector potential can be written as
\begin{eqnarray}
\mathcal{A}_{x\,{\rm tr}}(x,t)&=&-H_{\rm tr}^{(1)}\dfrac{k_xk_y}{k^3} \sin[k_x(x-D)-\omega t+\phi_{\rm tr}^{(1)}]
\notag\\
&&-H_{\rm tr}^{(2)}\dfrac{k_z}{k^2} \sin[k_x(x-D)-\omega t+\phi_{\rm tr}^{(2)}],
\notag\\
\mathcal{A}_{y\,{\rm tr}}(x,t)&=&-H_{\rm tr}^{(1)}\dfrac{k^2-k_y^2}{k^3}
\notag\\
&&\times\cos[k_x(x-D)-\omega t+\phi_{\rm tr}^{(1)}],
\label{A_tr}
\\
\mathcal{A}_{z\,{\rm tr}}(x,t)&=&H_{\rm tr}^{(1)}\dfrac{k_yk_z}{k^3} \cos[k_x(x-D)-\omega t+\phi_{\rm tr}^{(1)}]
\notag\\
&&-H_{\rm tr}^{(2)}\dfrac{k_x}{k^2} \cos[k_x(x-D)-\omega t+\phi_{\rm tr}^{(2)}].
\notag
\end{eqnarray}
where $H_{\rm tr}^{(1)}$, $\phi_{\rm tr}^{(1)}$, $H_{\rm tr}^{(2)}$, and $\phi_{\rm tr}^{(2)}$ are the amplitudes and phases of the magnetic field for the transmitted waves of the H$_\perp$ and E$_\perp$ polarizations.

\subsection{Electromagnetic field in the layered superconductor}

The electromagnetic field in the layered superconductor is defined by the distribution $\varphi({\vec r},t)$ of interlayer gauge-invariant phase difference of the order parameter. This phase difference is governed by a set of coupled sine-Gordon equations~\cite{sine-gord,SG2,SG3,SG4,SG5,Thz-rev}. Though this set does not take into account some important features of the cuprate high-temperature superconductors (e.g., the d-wave pairing), it describes properly the propagation of the electromagnetic waves in layered superconductors and allows important predictions.  For instance, a way to produce coherent terahertz radiation was proposed in  Ref.~\onlinecite{Bul_kosh} on the basis of the coupled sine-Gordon equations and then realized in the experiment~\cite{Ozyuzer}.

In the continuum limit, the coupled sine-Gordon equations reduce to
\begin{equation}\label{3}
\left(1-\lambda_{ab}^2\frac{\partial^2}{\partial
z^2}\right)\left(\frac{1}{\omega_J^2}\frac{\partial^2
\varphi}{\partial t^2} + \sin\varphi\right)-
\lambda_c^2\left(\frac{\partial^2 \varphi}{\partial x^2}+\frac{\partial^2 \varphi}{\partial y^2}\right)=0.
\end{equation}
Here $\lambda_{ab}$ and $\lambda_c=c/\omega_J\varepsilon^{1/2}$ are the London penetration depths across and along the layers, respectively, $\omega_J = (8\pi e d J_c/\hbar\varepsilon)^{1/2}$ is the Josephson plasma frequency, $J_c$ is the maximal Josephson current density, and $e$ is the elementary charge.  We do not take into account the relaxation terms since they are small at low temperatures and do not play an essential role in the phenomena considered here.

Note that the component $E_{z}$ of the electric field causes the breakdown of electro-neutrality of the superconducting layers and results in an additional, so-called capacitive, interlayer coupling. However, this coupling does not affect the propagation of the Josephson plasma waves along the waveguide and can be safely neglected because of the smallness of the parameter $\alpha = R_D^2\varepsilon/sd$. Here $R_D$ is the Debye length for a charge in the superconductor. In this case, the gauge of the vector potential can be chosen in such a way that the order parameter is real and the gauge-invariant phase difference $\varphi$ is related to the $z$-component of the vector potential by a simple expression (see, e.g., Ref.~\onlinecite{SG3}):
\begin{equation}\label{Az}
A_z = - \frac{\Phi_0}{2\pi d} \varphi,
\end{equation}
where $\Phi_0=\pi c \hbar/e$ is the magnetic flux quantum.

Note that Eq.~(\ref{3}) can be obtained from the more general wave equation,
\begin{equation}\label{wave_equation}
{\rm grad}\,{\rm div}{\vec A} - \Delta {\vec A}= -
\frac{\varepsilon}{c^2}\frac{\partial ^2 {\vec A}}{\partial t^2} +
\frac{4\pi}{c}{\vec J},
\end{equation}
with the current components
\begin{equation}\label{Jab}
J_x = - \frac{c}{4\pi\lambda_{ab}^2}A_x, \quad J_y = -
\frac{c}{4\pi\lambda_{ab}^2}A_y,
\end{equation}
and
\begin{equation}\label{Jc}
J_z = J_c \sin \varphi = - J_c \sin\left(\frac{2\pi d}{\Phi_0}A_z\right).
\end{equation}
Indeed, excluding  $A_x$ and $A_y$ from Eq.~(\ref{wave_equation}) and using Eqs.~(\ref{Jab}) and (\ref{Jc}), we derive Eq.~\eqref{3}.

Relations (\ref{HE}), (\ref{wave_equation}), (\ref{Jab}), and (\ref{Jc}) represent a complete set of equations for finding the electromagnetic field in the layered superconductor in the continual limit. We will use it for the study of the propagation of weakly nonlinear JPWs with $|\varphi|\ll 1$, when the density of the Josephson current can be presented as $J_c (\varphi-\varphi^3/6)$. It is important to emphasize that strongly nonlinear phenomena can be observed in the propagation of the JPWs even in the low-amplitude case, at  $|\varphi|\ll 1$, if the wave frequency is close to the cutoff frequency $\omega_{\rm cut}$ (see Refs.~\onlinecite{nl1,nl2,nl3,nl4}). Here $\omega_{\rm cut}$ is the minimum frequency of the linear JPWs that can propagate in the layered superconductor.

It is convenient to represent the electromagnetic field inside of a layered superconductor as a sum of waves with so-called \textit{ordinary} and \textit{extraordinary} polarizations. The electric field in the ordinary wave is perpendicular to the $z$-axis. Hence, the phase difference $\varphi$ and the component $A_z$ of the vector potential are equal to zero for this wave. Thus, the ordinary wave does not produce a current along the $z$-axis and, therefore, this mode is always \textit{linear}. Taking into account that $A_z=0$, one can solve Eq.~\eqref{wave_equation} and obtain components of the vector potential for the ordinary modes,
\begin{eqnarray}
\mathcal{A}_x^{\rm ord}&=&\dfrac{k_y}{k^2}\Big[e^{-p_xx}H_-^{\rm ord}\sin(\omega t-\phi_-^{\rm ord})
\notag\\
&&+\,e^{p_x(x-D)}H_+^{\rm ord}\sin(\omega t-\phi_+^{\rm ord})\Big],
\notag\\
\mathcal{A}_y^{\rm ord}&=&\dfrac{p_x}{k^2}\Big[e^{-p_xx}H_-^{\rm ord}\sin(\omega t-\phi_-^{\rm ord})
\notag\\
&&-\,e^{p_x(x-D)}H_+^{\rm ord}\sin(\omega t-\phi_+^{\rm ord})\Big],
\\
\mathcal{A}_z^{\rm ord}&=&0,
\notag
\end{eqnarray}
where $H_-^{\rm ord}$,  $\phi_-^{\rm ord}$, $H_+^{\rm ord}$, and $\phi_+^{\rm ord}$ are the amplitudes and phases of the decreasing and increasing evanescent fields inside the superconductor, $p_x=\lambda_{ab}^{-1}$.

The extraordinary polarization is perpendicular to the ordinary one, and the magnetic field in the wave of this polarization is perpendicular to the $z$-axis. This mode exhibits the \textit{nonlinear} properties of the Josephson plasma. We seek $A_z$ in the form of a wave with $x$-dependent amplitude $a(x)$ and phase $\eta(x)$,
\begin{equation}\label{varphi}
\mathcal{A}_z^{\rm ext}=
\mathcal{H}_0\tilde{\Omega}\lambda_c a(x)\sin[\omega t-\eta(x)]
\end{equation}
with
\begin{equation}\label{kappa_pm}
\mathcal{H}_0=\frac{4\sqrt{2}}{3\pi}\dfrac{\Phi_0}{d\lambda_c},
\quad
\tilde{\Omega}=|\Omega^2-\Omega_{\rm cut}^2|^{1/2},
\quad
\Omega=\frac{\omega}{\omega_J}.
\end{equation}
The cutoff frequency $\Omega_{\rm cut}$ is the minimum frequency of the linear extraordinary JPW which can propagate in the layered superconductor,
\begin{equation}\label{omega-cut}
\Omega_{\rm cut} = \Big(1+\dfrac{k_y^2 \lambda_c^2}{1+\lambda_{ab}k_z^2}\Big)^{1/2}.
\end{equation}

Introducing the dimensionless coordinate and the normalized length of the sample,
\begin{equation}\label{notations_ab}
\xi=\frac{x}{\lambda_c}\tilde{\Omega}\,, \qquad \delta=\frac{D}{\lambda_c}\tilde{\Omega}\,,
\end{equation}
and substituting Eq.~\eqref{varphi}  into Eq.~\eqref{wave_equation}, one can obtain the other components of the vector potential for the extraordinary modes:
\begin{align}
&\mathcal{A}_x^{\rm ext}=\mathcal{H}_0\tilde{\Omega}^2\lambda_{ab}^2 k_z\left[a\sin(\omega t-\eta)\right]^\prime,
\\\notag
&\mathcal{A}_y^{\rm ext}=\mathcal{H}_0\tilde{\Omega} \lambda_{ab}^2 \lambda_ck_yk_z a\sin(\omega t-\eta),
\end{align}
and two differential equations for the functions $\eta (\xi)$ and $a(\xi)$,
\begin{eqnarray}\label{from_sine-Gordon}
(a^2\eta^\prime)'=0,\quad
a^{\prime\prime}=- \sigma a-a^3+a{\eta^\prime}^{2}.
\end{eqnarray}
Here $\sigma = {\rm sign} (\Omega-\Omega_{\rm cut})$ and the prime denotes derivation over $\xi$. We will use these equations for the numerical calculations of the electromagnetic field distribution inside the sample of the layered superconductor.

\subsection{Superposition principle}

Matching the tangential components of the electric and magnetic fields at both interfaces (at $x=0$ and $x=D$) between the vacuum regions and the layered superconductor, we obtain two sets of equations. The boundary conditions at the interface $x=0$ give the equations,
\begin{subequations}\label{sys1}
\begin{eqnarray}
&\mu\big[\tilde{h}_{\rm inc}^{(1)}+\tilde{h}_{\rm ref}^{(1)}\big]=i(k_x\lambda_{c})^{-1}\alpha\gamma\tilde{h}_-^{\rm ord}
-i\gamma^{2} a(0) e^{i \eta(0)},
\label{sys1-1}\quad\\
&\tilde{h}_{\rm inc}^{(1)}+\tilde{h}_{\rm ref}^{(1)}-\alpha \big[\tilde{h}_{\rm inc}^{(2)}-\tilde{h}_{\rm ref}^{(2)}\big]=i a(0) e^{i \eta(0)},
\label{sys1-2}\quad\\
&\mu\big[\tilde{h}_{\rm inc}^{(2)}+\tilde{h}_{\rm ref}^{(2)}\big]=\gamma^{2}\tilde{h}_-^{\rm ord}-\beta \big[a(\xi)e^{i\eta(\xi)}\big]_{\xi=0}',
\label{sys1-3}\quad\\
&\alpha\big[\tilde{h}_{\rm inc}^{(1)}-\tilde{h}_{\rm ref}^{(1)}\big]+\tilde{h}_{\rm inc}^{(2)}+\tilde{h}_{\rm ref}^{(2)}=
-\tilde{h}_-^{\rm ord}.
\label{sys1-4}\quad
\end{eqnarray}
\end{subequations}
Here we introduce the normalized amplitudes of the waves of H$_\perp$ and E$_\perp$ polarizations in the vacuum and of the ordinary waves in the layered superconductor,
\begin{align}\label{h_nl}
&\tilde{h}_{\rm inc,\,ref,\,tr}^{(1),\,(2)}=h_{\rm inc,\,ref,\,tr}^{(1),\,(2)} \exp\big[{i\phi_{\rm inc,\,ref,\,tr}^{(1),\,(2)}}\big],
\notag\\
&h_{\rm inc,\,ref,\,tr}^{(1),\,(2)}=\dfrac{k_yk_z}{\mathcal{H}_0k^3\tilde{\Omega}\lambda_c}
H_{\rm inc,\,ref,\,tr}^{(1),\,(2)},\\
&\tilde{h}_{\pm}^{\rm ord}=h_{\pm}^{\rm ord}\exp({i\phi_{\pm}^{\rm ord}}),\quad
h_{\pm}^{\rm ord}=\dfrac{H_{\pm}^{\rm ord}}{\mathcal{H}_0k^3\tilde{\Omega}\lambda_c\lambda_{ab}^2},\notag
\end{align}
and the parameters,
\begin{gather}
\label{const}
\alpha=\dfrac{kk_x}{k_yk_z},
\;
\beta=\dfrac{\tilde{\Omega}}{kk_yk_z\lambda_{c}^{3}},
\;
\mu=\dfrac{k^2-k_y^2}{k_y^{2}k_z^{2}\lambda_{c}^{2}},
\;
\gamma=\dfrac{\lambda_{ab}}{\lambda_c}.
\end{gather}

The boundary conditions at the interface $x=D$ give the equations,
\begin{subequations}\label{sys2}
\begin{eqnarray}
&\mu\tilde{h}_{\rm tr}^{(1)}=-i(k_x\lambda_{c})^{-1}\alpha\gamma\tilde{h}_+^{\rm ord}-i\gamma^{2} a(\delta) e^{i\eta(\delta)},
\label{sys2-1}\\
&\tilde{h}_{\rm tr}^{(1)}-\alpha \tilde{h}_{\rm tr}^{(2)}=i a(\delta) e^{i\eta(\delta)},
\label{sys2-2}\\
&\mu\tilde{h}_{\rm tr}^{(2)}=\gamma^{2}\tilde{h}_+^{\rm ord}-\beta\big[a(\xi)e^{i\eta(\xi)}\big]_{\xi=\delta}',
\label{sys2-3}\\
&\alpha\tilde{h}_{\rm tr}^{(1)}+\tilde{h}_{\rm tr}^{(2)}=-\tilde{h}_+^{\rm ord}.\label{sys2-4}
\end{eqnarray}
\end{subequations}
Here we omit the terms with $\exp(-p_xD)$ because we assume that the sample length $D$ is much larger than the London penetration depth $\lambda_{ab}$.

Since $\gamma=\lambda_{ab}/\lambda_c\ll1$, the right-hand side in Eq.~\eqref{sys1-1} is relatively small. So, in the main approximation with respect to the anisotropy parameter $\gamma$, we have
\begin{equation}
\tilde{h}_{\rm ref}^{(1)}\approx
-\tilde{h}_{\rm inc}^{(1)}.
\end{equation}
Correspondingly, Eq.~\eqref{sys2-1} shows that the amplitude $\tilde{h}_{\rm tr}^{(1)}$ of the transmitted wave of the H$_\perp$ polarization is much less than the incident amplitude $\tilde{h}_{\rm inc}^{(1)}$,  $|\tilde{h}_{\rm tr}^{(1)}/\tilde{h}_{\rm inc}^{(1)}|\sim \gamma\ll1$. This means that the incident mode of the H$_\perp$ polarization \textit{nearly completely reflects} from the layered superconductor. Moreover, this behavior of the H$_\perp$-polarized wave \textit{does not depend on the presence of the orthogonal mode with the E$_\perp$ polarization}.

In the main approximation with respect to $\gamma$, relations~\eqref{sys1-2}, \eqref{sys1-3},~\eqref{sys2-2}, and~\eqref{sys2-3} give the following equations, which describe the reflection and transmission of the wave with the E$_\perp$ polarization:
\begin{subequations}\label{sys3}
\begin{eqnarray}
&-\alpha \big(\tilde{h}_{\rm inc}^{(2)}-\tilde{h}_{\rm ref}^{(2)}\big)=i a(0) e^{i \eta(0)},
\\
&\mu\big(\tilde{h}_{\rm inc}^{(2)}+\tilde{h}_{\rm ref}^{(2)}\big)=-\beta e^{i\eta(0)}\left[a^\prime(0)+ia(0)\eta^\prime(0)\right],
\\
&-\alpha \tilde{h}_{\rm tr}^{(2)}=i a(\delta) e^{i\eta(\delta)},
\\
&\mu\tilde{h}_{\rm tr}^{(2)}=-\beta e^{i\eta(\delta)}\big[a^\prime(\delta)+i a(\delta)\eta^\prime(\delta)\big].
\end{eqnarray}
\end{subequations}
This set of equations, as well as Eq.~\eqref{from_sine-Gordon}, do not contain the parameters of the H$_\perp$-polarized wave. This means that the reflection and transmission of the wave with the E$_\perp$ polarization \textit{do not depend on the presence of the mode of the H$_\perp$ polarization}. Thus, we have shown that, in the main approximation with respect to the anisotropy parameter $\gamma$, the modes of the H$_\perp$ and E$_\perp$ polarizations reflect and transmit through the layered superconductor independently of each other. This means that the superposition principle for these specific modes works even in the strongly-nonlinear regime.

After calculating the reflected and transmitted amplitudes for the modes of the H$_\perp$ and E$_\perp$ polarizations, we can use Eqs.~\eqref{sys1-4} and~\eqref{sys2-4} to determine the amplitudes $\tilde{h}_-^{\rm ord}$ and $\tilde{h}_+^{\rm ord}$ of the ordinary modes in the layered superconductor.

\begin{figure}
\begin{center}
\includegraphics*[width=14cm]{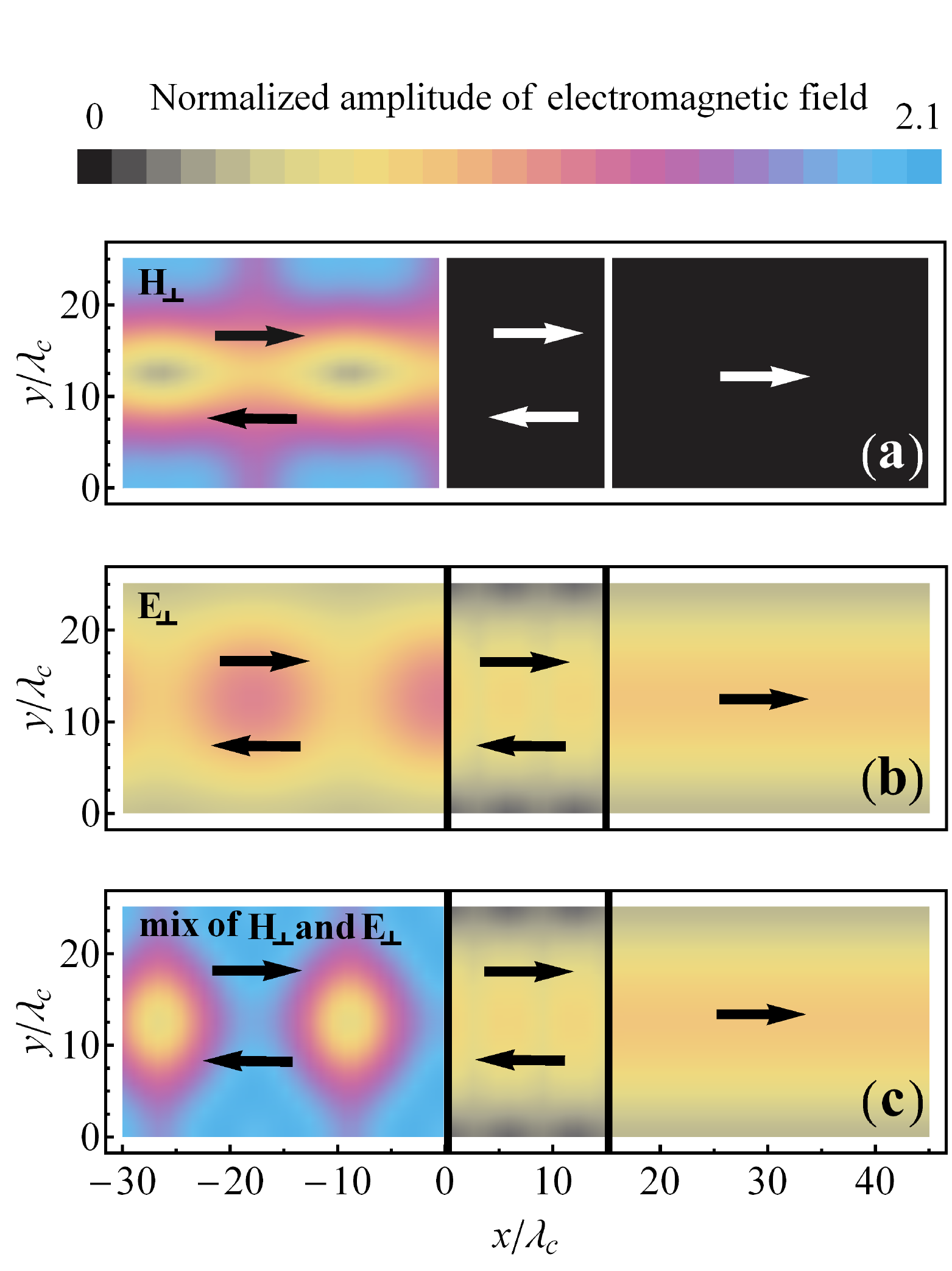}
\caption{(Color online) Spatial distribution (over coordinates $x$ and $y$ at $z=L_z/3$) of the normalized amplitude $\bar{W}$ of electromagnetic field, Eq.~\eqref{norm-EMF}, inside the waveguide. Panels (a), (b), and (c) correspond to the cases when the incident wave is of the H$_\perp$ polarization only, of the E$_\perp$ polarization only, and a mix of the H$_\perp$ and E$_\perp$ polarizations, respectively. In the panel (a): $h_{\rm inc}^{(1)}=2$, $h_{\rm inc}^{(2)}=0$; in the panel (b): $h_{\rm inc}^{(1)}=0$, $h_{\rm inc}^{(2)}=2$; in the panel (c): $h_{\rm inc}^{(1)}=h_{\rm inc}^{(2)}=2$. The color determines the value of the amplitude. The vertical straight lines show the edges of the superconducting sample. The parameters used here are: $\tilde{\Omega}= 0.1$, $\sigma=-1$, $D=15\lambda_c$, $L_y=L_z=0.1$~cm, $n_y=n_z=1$, $\phi_{\rm inc}^{(1)}=\phi_{\rm inc}^{(2)}=0$, $\lambda_c=4\cdot 10^{-3}$~cm, $\lambda_{ab}=2000$~\AA, $\omega_J/2\pi=0.3$~THz.
}
\label{colorplots}
\end{center}
\end{figure}

In order to illustrate the superposition principle, we solve three problems of reflection and transmission of the incident waves: the incident mode has only H$_\perp$ polarization only, with $h_{\rm inc}^{(1)}=2$; the incident mode is of the E$_\perp$ polarization only with $h_{\rm inc}^{(2)}=2$; the incident mode is a superposition of the waves of H$_\perp$ and E$_\perp$ polarizations with $h_{\rm inc}^{(1)}=h_{\rm inc}^{(2)}=2$. Figure~\ref{colorplots} shows the distribution of the normalized characteristic amplitude $\bar{W}$ of the electromagnetic field inside of the waveguide for these three cases,
\begin{equation}
\label{norm-EMF}
\bar{W}=\dfrac{k_yk_z}{\mathcal{H}_0k^3\tilde{\Omega}\lambda_c}
\sqrt{\big\langle|{\vec E}(\vec{r},t)|^{2}+|{\vec H}(\vec{r},t)|^{2}\big\rangle_t},
\end{equation}
where $\langle\ldots\rangle_t$ denotes averaging over time $t$.

As seen in Fig.~\ref{colorplots}(a), the wave of H$_\perp$ polarization completely reflects from the sample of the layered superconductor and does not excite a propagating mode inside it. Only the evanescent mode, which decays in a very narrow region near the surface of the sample, exists. Hence, there is also no transmitted wave. The wave of E$_\perp$ polarization, Fig.~\ref{colorplots}(b), partly reflects from the layered superconductor and partly transmits through it. When we apply the mix of the waves of the H$_\perp$ and E$_\perp$ polarizations with the same amplitudes, Fig.~\ref{colorplots}(c), the field distribution in the superconductor and in the right vacuum region is the same as in Fig.~\ref{colorplots}(b). This demonstrates that the modes of the H$_\perp$ and E$_\perp$ polarizations do not interact with each other and can be treated independently.

\section{Transmission and reflection of the E$_\perp$-polarized waves}\label{Ep}

In this section, we consider the nonlinear phenomena in the reflection and transmission of the E$_\perp$-polarized waves through the sample of the layered superconductor placed inside of the waveguide. Using Eqs.~\eqref{from_sine-Gordon} and~\eqref{sys3}, we calculate the amplitudes of the reflected and transmitted waves. We start from the analysis of the phase trajectories $a'(a)$ that correspond to certain solutions of these equations. Excluding $\eta'$ from Eqs.~\eqref{from_sine-Gordon} and~\eqref{sys3}, we derive the explicit equations for the phase trajectories,
\begin{eqnarray}
\label{phase-tr}
{a'}^{2}(a)
&=&\sigma\big[a^{2}(\delta)-a^{2}\big]
+\dfrac{1}{2}\big[a^{4}(\delta)-a^{4}\big]
\notag\\
&&+\dfrac{a^{4}(\delta)}{(\alpha\beta/\mu)^{2}}
\big[a^{-2}(\delta)-a^{-2}\big].
\end{eqnarray}

\begin{figure}
\begin{center}
\includegraphics*[width=14cm]{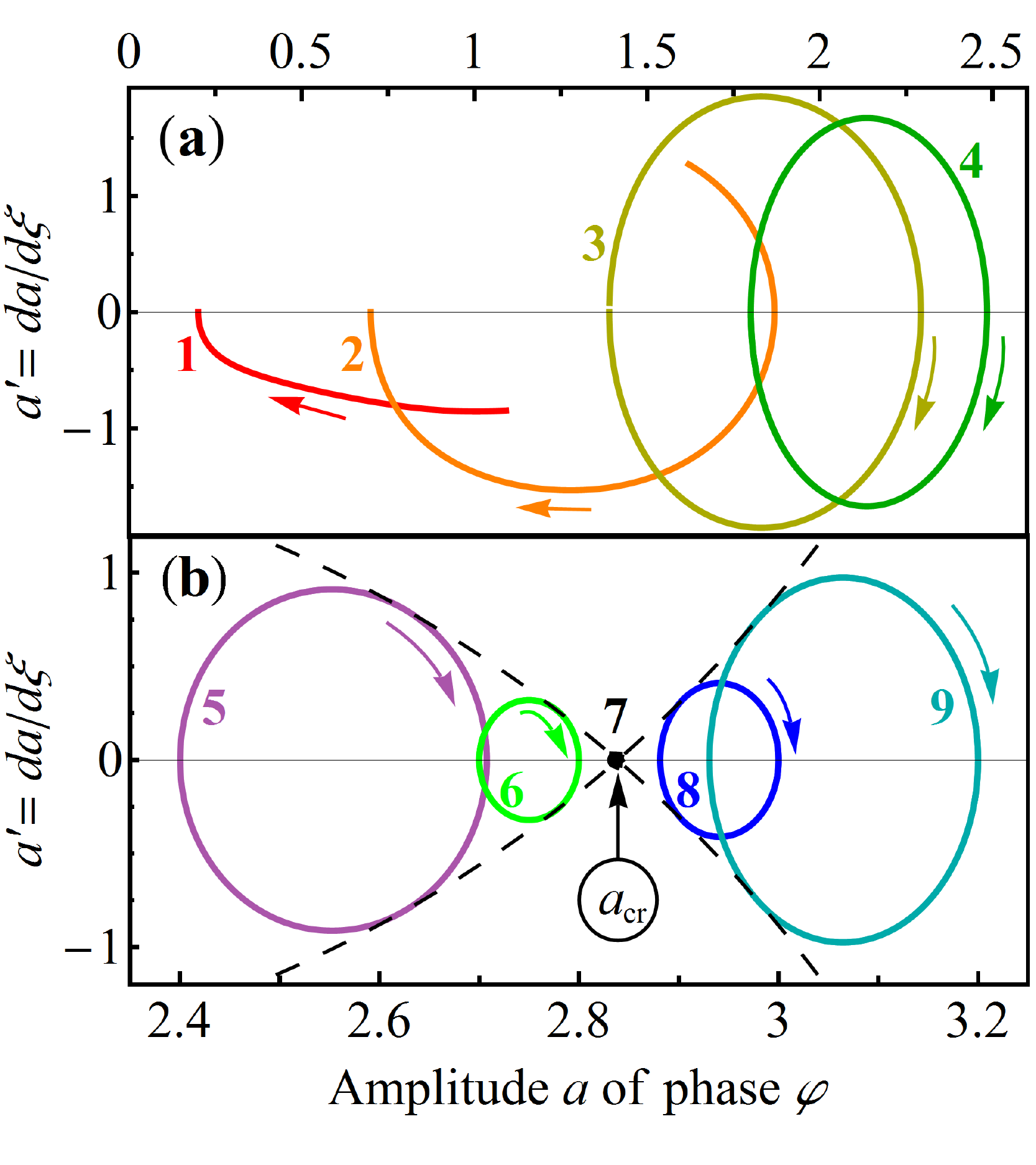}
\caption{(Color online)
Phase trajectories $a'(a)$ plotted for $\delta=3$ ($D=30\lambda_c$) that correspond to the solid thick blue curve in Fig.~\ref{T(hiII)}. Other parameters are the same as in Fig.~\ref{colorplots}. (a) Unclosed phase trajectories plotted for $a(\delta)$=0.2 (red curve 1) and $a(\delta)$=0.7 (orange curve 2); nearly closed loop plotted for $a(\delta)$=1.39 (brown curve 3); the green curve 4 plotted for $a(\delta)$=1.8 is a loop with overlapping portions. The movement along the phase trajectory, when the spatial coordinate $\xi$ changes from zero to $\delta$, is shown by the arrows. (b) Phase trajectories plotted for $a(\delta)$=2.4 (purple curve 5), $a(\delta)$=2.7 (green curve 6), $a(\delta)$=3 (blue curve 8), and $a(\delta)$=3.2 (dark cyan curve 9). The black dashed curves are the envelopes for the phase trajectories. The point 7 is a shrunken phase trajectory plotted for $a(\delta) = a_{\rm cr} \approx 2.84$. }
\label{phase_diagr}
\end{center}
\end{figure}

Figure~\ref{phase_diagr} presents the phase diagram of Eqs.~\eqref{phase-tr} for different values of the integration constant $a(\delta)$. The arrows show the direction of movement when increasing the coordinate $\xi$. This figure demonstrates how the phase trajectories $a'(a)$ change when increasing $a(\delta)$. The phase trajectories are open loops at $a(\delta) < a_1 $ (e.g., curves 1 and 2 in Fig.~\ref{phase_diagr}). For the set of parameters considered in Fig.~\ref{phase_diagr}, $a_1 \approx 1.39$. If $a(\delta) > a_1$, the phase trajectories have overlapping portions (curves 4--9 in Fig.~\ref{phase_diagr}). We stress that there exists a special value of $a(\delta) = a_{\rm cr} \approx 2.84$, where the phase trajectory $a'(a)$ shrinks into a point (see point 7 in Fig.~\ref{phase_diagr}. This point corresponds to the uniform spatial distribution of the amplitude $a(\xi)$, i.e. $a(\xi)=a_{\rm cr}$, for all $\xi$.

The uniform solution $a(\xi)=a_{\rm cr}$ occurs when the amplitude $H_{\rm inc}^{(2)}$ of the incident wave takes on a critical value $H_{\rm cr}$,
\begin{equation}
\label{hcr}
H_{\rm cr}=\mathcal{H}_0\dfrac{k^2\lambda_c}{k_x}
\sqrt{\Big[\dfrac{(k^2-k_y^2) \lambda_c}{k_x}\Big]^{2}+ \dfrac{\omega_{\rm cut}^2-\omega^2}{\omega_J^{2}}}.
\end{equation}
In this case, according to Eqs.~\eqref{from_sine-Gordon}, the phase $\eta$ changes linearly with $\xi$. This means that the electromagnetic field inside the sample behaves similarly to a linear wave propagating only in one direction (along the $x$-axis in Fig.~\ref{wavegAB}), and there is no wave reflected from the boundary $x=D$. The superconducting slab is totally transparent in this case. Note that the phase trajectories for $a<a_{\rm cr}$ and $a>a_{\rm cr}$ correspond to the amplitudes $H_{\rm inc}^{(2)}<H_{\rm cr}$ and $H_{\rm inc}^{(2)}>H_{\rm cr}$, respectively.

\begin{figure}
\begin{center}
\includegraphics*[width=12cm]{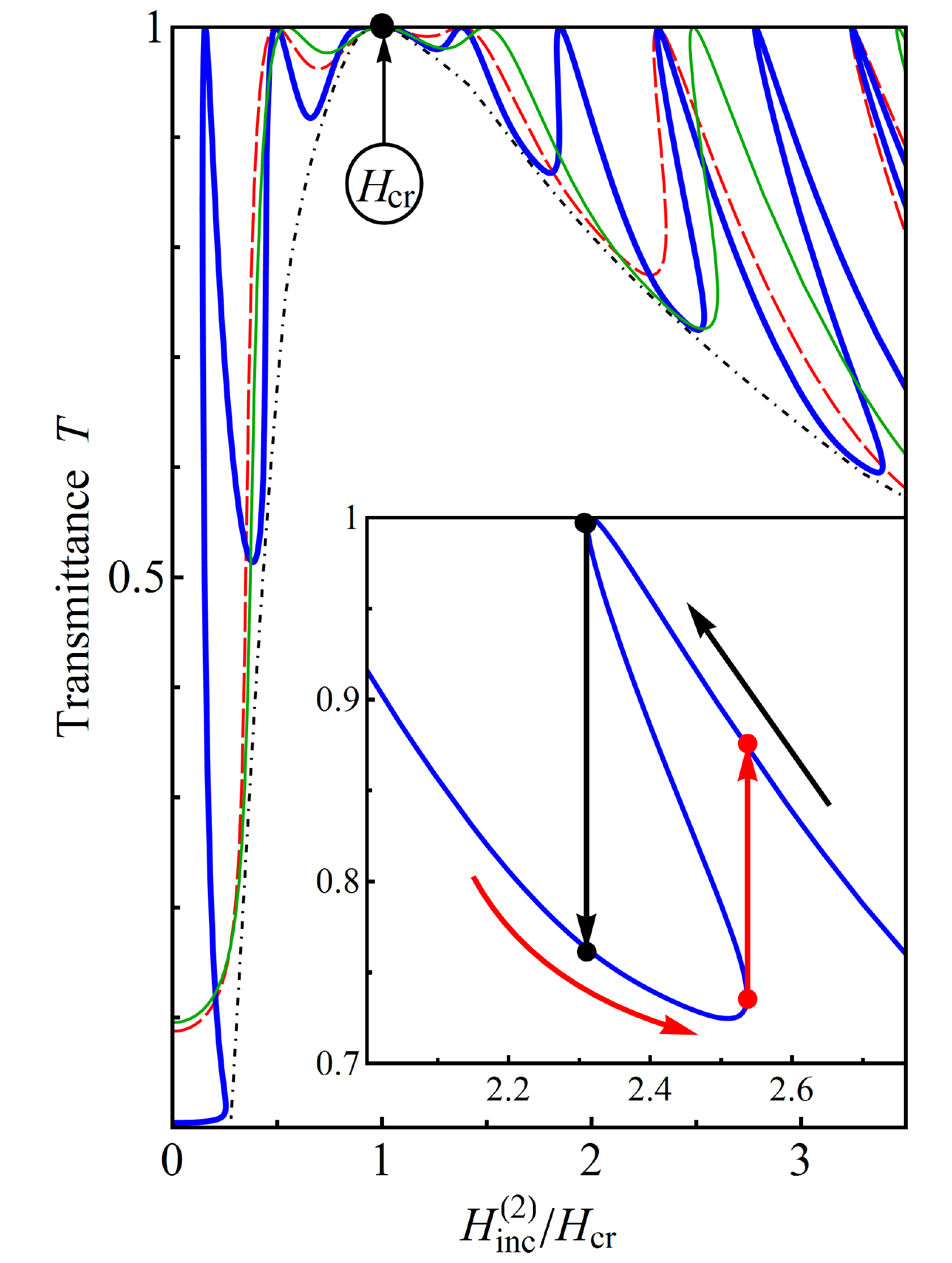}
\caption{(Color online) Transmittance $T$ versus the normalized amplitude $H_{\rm inc}^{(2)}/H_{\rm cr}$ (see Eq.~\eqref{hcr}) of the incident E$_\perp$-polarized wave.
\\
\textbf{Main panel.} Curves $T(H_{\rm inc}^{(2)}/H_{\rm cr})$ plotted for different sizes of the sample. Solid thick blue curve corresponds to $L_y=L_z=0.1$~cm, $D=30\lambda_c$ ($\delta=3$); solid thin green curve is plotted for $L_y=L_z=0.3$~cm, $D=15\lambda_c$ ($\delta=1.5$); dashed thin red curve corresponds to $L_y=L_z=0.1$~cm, $D=15\lambda_c$ ($\delta=1.5$). Other parameters are the same as in Fig.~\ref{colorplots}. The black dash-dotted curve is the envelope for all $T(H_{\rm inc}^{(2)}/H_{\rm cr})$ curves.
\\
\textbf{Inset.} Hysteresis of the $T(H_{\rm inc}^{(2)}/H_{\rm cr})$ dependence when moving along  the solid thick blue curve shown in the main panel. The red lower and upward arrows show movement along the $T(H_{\rm inc}^{(2)}/H_{\rm cr})$ curve when increasing amplitude $H_{\rm inc}^{(2)}$. The black upper and downward arrows correspond to decreasing amplitude $H_{\rm inc}^{(2)}$.}
\label{T(hiII)}
\end{center}
\end{figure}

Figure~\ref{T(hiII)} shows the dependence of the transmittance $T$ on the amplitude $H_{\rm inc}^{(2)}$ (normalized to $H_{\rm cr}$) of the incident E$_\perp$-polarized wave for different sizes of $D$, $L_x$, and $L_y$ of the superconducting sample.
All the curves in the main panel of Fig.~\ref{T(hiII)} have an oscillating structure so that the transmittance $T$ tops the maximum value $T=1$ at different amplitudes of the incident wave. Besides the considered case $H_{\rm inc}^{(2)}=H_{\rm cr}$, the complete transmission of the sample ($T=1$) occurs in conditions when the phase trajectories $a'(a)$ in Fig.~\ref{phase_diagr} represent closed loops with integer numbers of complete turns along them, i.e., for all cases when $a(0)=a(\delta)$. Physically, such conditions correspond to the cases when the sample length $D$ is equal to the integer number of the wavelengths of the nonlinear mode.

All the curves $T(H_{\rm inc}^{(2)}/H_{\rm cr})$ in Fig.~\ref{T(hiII)} plotted for samples with different sizes $L_x$, $L_y$, and $D$ have the common envelope curve (the dash-dotted curve). The critical amplitude $H_{\rm inc}^{(2)}=H_{\rm cr}$ is the only point, where all the curves and the envelope curve touch each other. It should be noted that the $T(H_{\rm inc}^{(2)}/H_{\rm cr})$ dependence plotted for another frequency detuning ${(\omega-\omega_{\rm cut})}$ behaves similarly to the curves shown in the main panel in Fig.~\ref{T(hiII)}, however with another envelope.

The dependence $T(H_{\rm inc}^{(2)}/H_{\rm cr})$ has interesting hysteretic features. The inset in Fig.~\ref{T(hiII)} shows a fragment of such a dependence. When increasing the amplitude $H_{\rm inc}^{(2)}$, one should move along the red bottom arrow. When the right endpoint on this branch is reached, further movement along this branch is not possible. Increasing the incident wave amplitude results in a jump to the higher branch, along the red upward arrow. Similar jump occurs when decreasing the amplitude $H_{\rm inc}^{(2)}$, see the black upper and downward arrows. At first the transmittance increases, but when the left endpoint on this branch is reached, a jump to the lower branch takes place.

\section{Transmission and reflection of the TE and TM modes}

In this section, we study the nonlinear transmission, reflection, and mutual transformation of the TE and TM modes in the waveguide with a sample of layered superconductor. For the geometry shown in Fig.~\ref{wavegAB}, the TE (TM) wave is defined as a mode with the electric (magnetic) field perpendicular to the $x$-axis. The vector potential $\vec{\mathcal{A}}_{\rm inc}^{\rm (TE)}$ of the incident TE-polarized wave has the components,
\begin{eqnarray}\label{TE-fields_inc}
\mathcal{A}_{x\, {\rm inc}}^{\rm (TE)}&=&0,
\notag\\
\mathcal{A}_{y\, {\rm inc}}^{\rm (TE)}&=&-H_{\rm inc}^{\rm (TE)}\dfrac{kk_z}{k^3} \cos[k_xx-\omega t+\phi_{\rm inc}^{\rm (TE)}],
\\\notag
\mathcal{A}_{z\, {\rm inc}}^{\rm (TE)}&=&H_{\rm inc}^{\rm (TE)}\dfrac{kk_y}{k^3} \cos[k_xx-\omega t+\phi_{\rm inc}^{\rm (TE)}],
\end{eqnarray}
where $H_{\rm inc}^{\rm (TE)}$ and $\phi_{\rm inc}^{\rm (TE)}$ are the amplitude and phase of the magnetic field in this wave. For the incident TM-polarized wave with amplitude $H_{\rm inc}^{\rm (TM)}$ and phase $\phi_{\rm inc}^{\rm (TM)}$ of the magnetic field, we have
\begin{eqnarray}\label{TM-fields_inc}
\mathcal{A}_{x\, {\rm inc}}^{\rm (TM)}&=&H_{\rm inc}^{\rm (TM)}\dfrac{k^2-k_x^2}{k^3} \sin[k_xx-\omega t+\phi_{\rm inc}^{\rm (TM)}],
\notag\\
\mathcal{A}_{y\, {\rm inc}}^{\rm (TM)}&=&H_{\rm inc}^{\rm (TM)}\dfrac{k_xk_y}{k^3} \cos[k_xx-\omega t+\phi_{\rm inc}^{\rm (TM)}],\notag
\\
\mathcal{A}_{z\, {\rm inc}}^{\rm (TM)}&=&H_{\rm inc}^{\rm (TM)}\dfrac{k_xk_z}{k^3} \cos[k_xx-\omega t+\phi_{\rm inc}^{\rm (TM)}].
\end{eqnarray}

The components of the vector potential for the reflected TE and TM modes can be written in a similar form,
\begin{eqnarray}\label{TE-fields_ref}
\mathcal{A}_{x\, {\rm ref}}^{\rm (TE)}&=&0,
\notag\\
\mathcal{A}_{y\, {\rm ref}}^{\rm (TE)}&=&-H_{\rm ref}^{\rm (TE)}\dfrac{kk_z}{k^3} \cos[k_xx+\omega t-\phi_{\rm ref}^{\rm (TE)}],
\\\notag
\mathcal{A}_{z\, {\rm ref}}^{\rm (TE)}&=&H_{\rm ref}^{\rm (TE)}\dfrac{kk_y}{k^3} \cos[k_xx+\omega t-\phi_{\rm ref}^{\rm (TE)}],
\end{eqnarray}
\begin{eqnarray}\label{TM-fields_ref}
\mathcal{A}_{x\, {\rm ref}}^{\rm (TM)}&=&-H_{\rm ref}^{\rm (TM)}\dfrac{k^2-k_x^2}{k^3} \sin[k_xx+\omega t-\phi_{\rm ref}^{\rm (TM)}],
\notag\\
\mathcal{A}_{y\, {\rm ref}}^{\rm (TM)}&=&-H_{\rm ref}^{\rm (TM)}\dfrac{k_xk_y}{k^3} \cos[k_xx+\omega t-\phi_{\rm ref}^{\rm (TM)}],
\notag\\
\mathcal{A}_{z\, {\rm ref}}^{\rm (TM)}&=&-H_{\rm ref}^{\rm (TM)}\dfrac{k_xk_z}{k^3} \cos[k_xx+\omega t-\phi_{\rm ref}^{\rm (TM)}].
\end{eqnarray}

Finally, in the second vacuum region (for $x>D$), only the transmitted wave exists. In this region, the components of vector potential can be written as
\begin{eqnarray}\label{TE-fields_tr}
\mathcal{A}_{x\, {\rm tr}}^{\rm (TE)}&=&0,
\notag\\
\mathcal{A}_{y\, {\rm tr}}^{\rm (TE)}&=&-H_{\rm tr}^{\rm (TE)}\dfrac{kk_z}{k^3} \cos[k_x(x-D)-\omega t+\phi_{\rm tr}^{\rm (TE)}],
\\\notag
\mathcal{A}_{z\, {\rm tr}}^{\rm (TE)}&=&H_{\rm tr}^{\rm (TE)}\dfrac{kk_y}{k^3} \cos[k_x(x-D)-\omega t+\phi_{\rm tr}^{\rm (TE)}],
\end{eqnarray}
\begin{eqnarray}\label{TM-fields_tr}
\mathcal{A}_{x\, {\rm tr}}^{\rm (TM)}&=&H_{\rm tr}^{\rm (TM)}\dfrac{k^2-k_x^2}{k^3} \sin[k_x(x-D)-\omega t+\phi_{\rm tr}^{\rm (TM)}],
\notag\\
\mathcal{A}_{y\, {\rm tr}}^{\rm (TM)}&=&H_{\rm tr}^{\rm (TM)}\dfrac{k_xk_y}{k^3} \cos[k_x(x-D)-\omega t+\phi_{\rm tr}^{\rm (TM)}],
\notag\\
\mathcal{A}_{z\, {\rm tr}}^{\rm (TM)}&=&H_{\rm tr}^{\rm (TM)}\dfrac{k_xk_z}{k^3} \cos[k_x(x-D)-\omega t+\phi_{\rm tr}^{\rm (TM)}].
\end{eqnarray}

Evidently, the electromagnetic field of the TE and TM waves can be presented as superpositions of the H$_\perp$- and E$_\perp$-polarized waves. The analysis of Eqs.~\eqref{A_inc}--\eqref{A_tr} and~\eqref{TE-fields_inc}--\eqref{TM-fields_tr} gives the following expressions for the complex dimensionless amplitudes $\tilde{h}_{\rm inc, \, ref, \, tr}^{\rm (TE),\, (TM)}$ of the incident, transmitted, and reflected TE and TM waves via the amplitudes $\tilde{h}_{\rm inc,\,ref,\,tr}^{(1),(2)}$ of the H$_\perp$- and E$_\perp$-polarized waves:
\begin{eqnarray}\label{TE-TM_inc_tr}
\tilde{h}_{\rm inc, \, tr}^{\rm (TE)}&=&\dfrac{kk_z\tilde{h}_{\rm inc,\,tr}^{(1)}-k_xk_y\tilde{h}_{\rm inc,\,tr}^{(2)}}{k_y^2+k_z^2},
\notag\\
\quad
\tilde{h}_{\rm inc, \, tr}^{\rm (TM)}
&=&
-\dfrac{k_xk_y\tilde{h}_{\rm inc,\,tr}^{(1)}+kk_z\tilde{h}_{\rm inc,\,tr}^{(2)}}{k_y^2+k_z^2},
\end{eqnarray}
\begin{eqnarray}\label{TE-TM_ref}
\tilde{h}_{\rm ref}^{\rm (TE)}&=&\dfrac{kk_z\tilde{h}_{\rm ref}^{(1)}+k_xk_y\tilde{h}_{\rm ref}^{(2)}}{k_y^2+k_z^2},
\notag\\
\quad
\tilde{h}_{\rm ref}^{\rm (TM)}
&=&
\dfrac{k_xk_y\tilde{h}_{\rm ref}^{(1)}-kk_z\tilde{h}_{\rm ref}^{(2)}}{k_y^2+k_z^2}.
\end{eqnarray}
Here $\tilde{h}_{\rm inc,\,tr,\,ref}^{\rm (TE),\,(TM)}$ is defined similarly to Eq.~\eqref{h_nl},
\begin{eqnarray}\label{TE-12}
\tilde{h}_{\rm inc,\,tr,\,ref}^{\rm (TE),\,(TM)}&=&h_{\rm inc,\,tr,\,ref}^{\rm (TE),\,(TM)} \exp\big[{i\phi_{\rm inc,\,tr,\,ref}^{\rm (TE),\,(TM)}}\big],
\notag\\
h_{\rm inc,\,tr,\,ref}^{\rm (TE),\,(TM)}&=&\dfrac{k_yk_z}{\mathcal{H}_0k^3\tilde{\Omega}\lambda_c}
H_{\rm inc,\,tr,\,ref}^{\rm (TE),\,(TM)}.
\end{eqnarray}

\begin{figure}
\begin{center}
\includegraphics*[width=14cm]{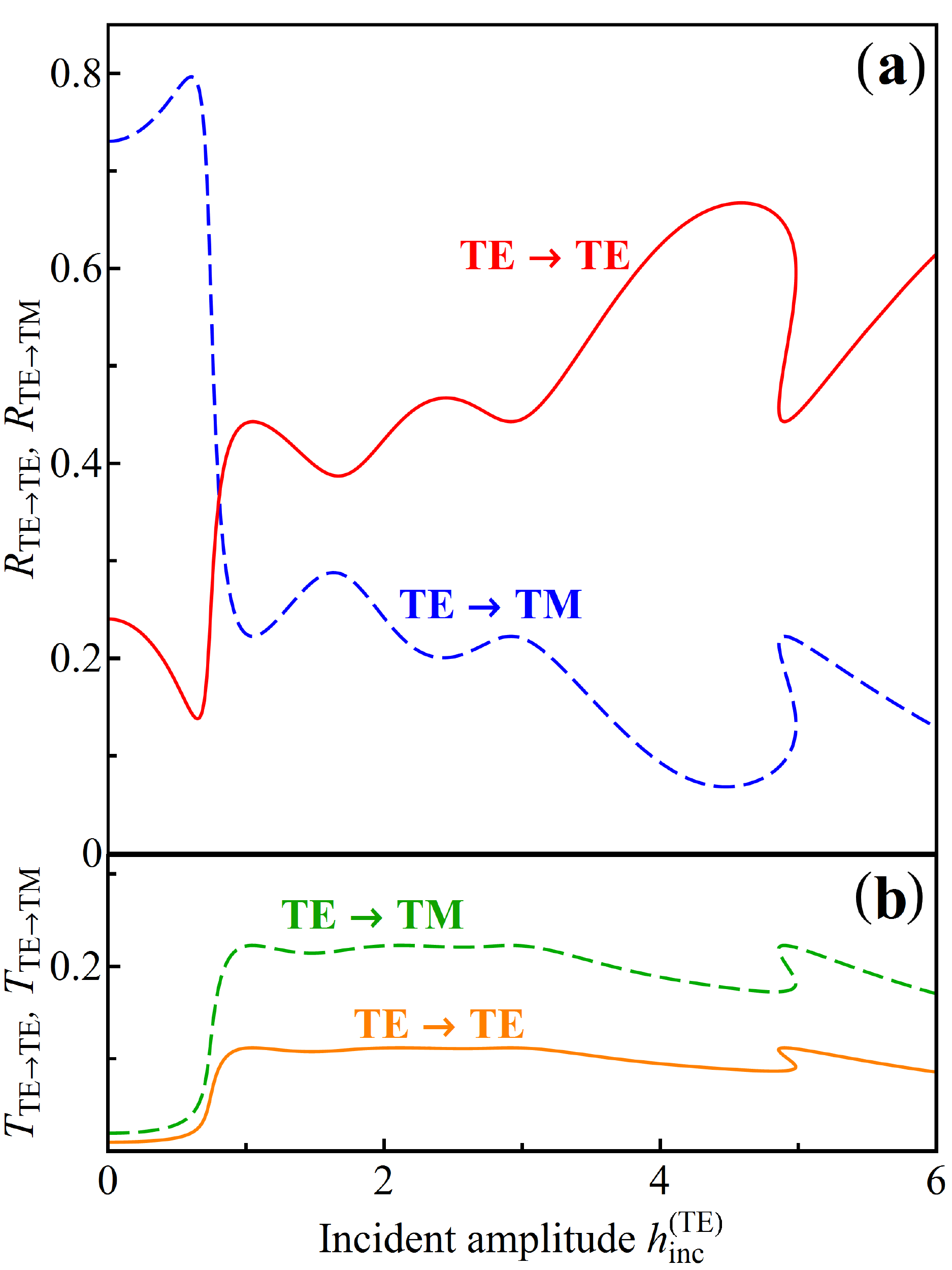}
\caption{(Color online) (a) Reflectance $R_{\rm TE\rightarrow TE}$ (red solid line) and transformation coefficient $R_{{\rm TE \rightarrow TM}}$ (blue dashed line) versus the dimensionless amplitude $h_{\rm inc}^{\rm (TE)}$ of the incident TE wave. (b) Transmittance $T_{\rm TE\rightarrow TE}$ (orange solid line) and transformation coefficient $T_{{\rm TE \rightarrow TM}}$ (green dashed line) versus the dimensionless amplitude $h_{\rm inc}^{\rm (TE)}$ of the incident TE wave. The parameters used here are the same as in Fig.~\ref{colorplots}.}
\label{TE}
\end{center}
\end{figure}

First, we consider the case when the wave incident onto the layered superconductor is TE-polarized. Using Eqs.~(\ref{TE-TM_inc_tr}) and (\ref{TE-TM_ref}), the superposition principle for the waves with the H$_\perp$ and E$_\perp$ polarizations, and the results of the previous sections on the nonlinear reflection and transmission of the H$_\perp$- and E$_\perp$-polarized waves, we can find the reflectance $R_{\rm TE\rightarrow TE}$ and transmittance $T_{\rm TE\rightarrow TE}$ for the TE waves,
\begin{equation}\label{R}
R_{{\rm TE\rightarrow TE}}=\Big|\dfrac{h_{\rm ref}^{\rm (TE)}}{h_{\rm inc}^{\rm (TE)}}\Big|^{2},
\quad
T_{{\rm TE\rightarrow TE}}=\Big|\dfrac{h_{\rm tr}^{\rm (TE)}}{h_{\rm inc}^{\rm (TE)}}\Big|^{2}.
\end{equation}
In addition, we obtain the transformation coefficients $R_{{\rm TE \rightarrow TM}}$ and $T_{{\rm TE \rightarrow TM}}$ for the TM waves that appear in the vacuum regions $x<0$ and $x>D$, respectively, due to the anisotropy in the $yz$-plane,
\begin{equation}\label{trans}
R_{{\rm TE \rightarrow TM}}=\Big|\dfrac{h_{\rm ref}^{\rm (TM)}}{h_{\rm inc}^{\rm (TE)}}\Big|^{2},
\quad
T_{{\rm TE \rightarrow TM}}=\Big|\dfrac{h_{\rm tr}^{\rm (TM)}}{h_{\rm inc}^{\rm (TE)}}\Big|^{2}.
\end{equation}
Figure~\ref{TE} shows the numerically-calculated dependences of the coefficients $R_{\rm TE\rightarrow TE}$, $T_{\rm TE\rightarrow TE}$, $R_{{\rm TE \rightarrow TM}}$, and $T_{{\rm TE \rightarrow TM}}$ on the dimensionless amplitude $h_{\rm inc}^{\rm (TE)}$ of the incident TE wave. Note that all these dependences exhibit hysteretic behavior when changing the incident amplitude, see the discussion in the previous section.


\begin{figure}
\begin{center}
\includegraphics*[width=14cm]{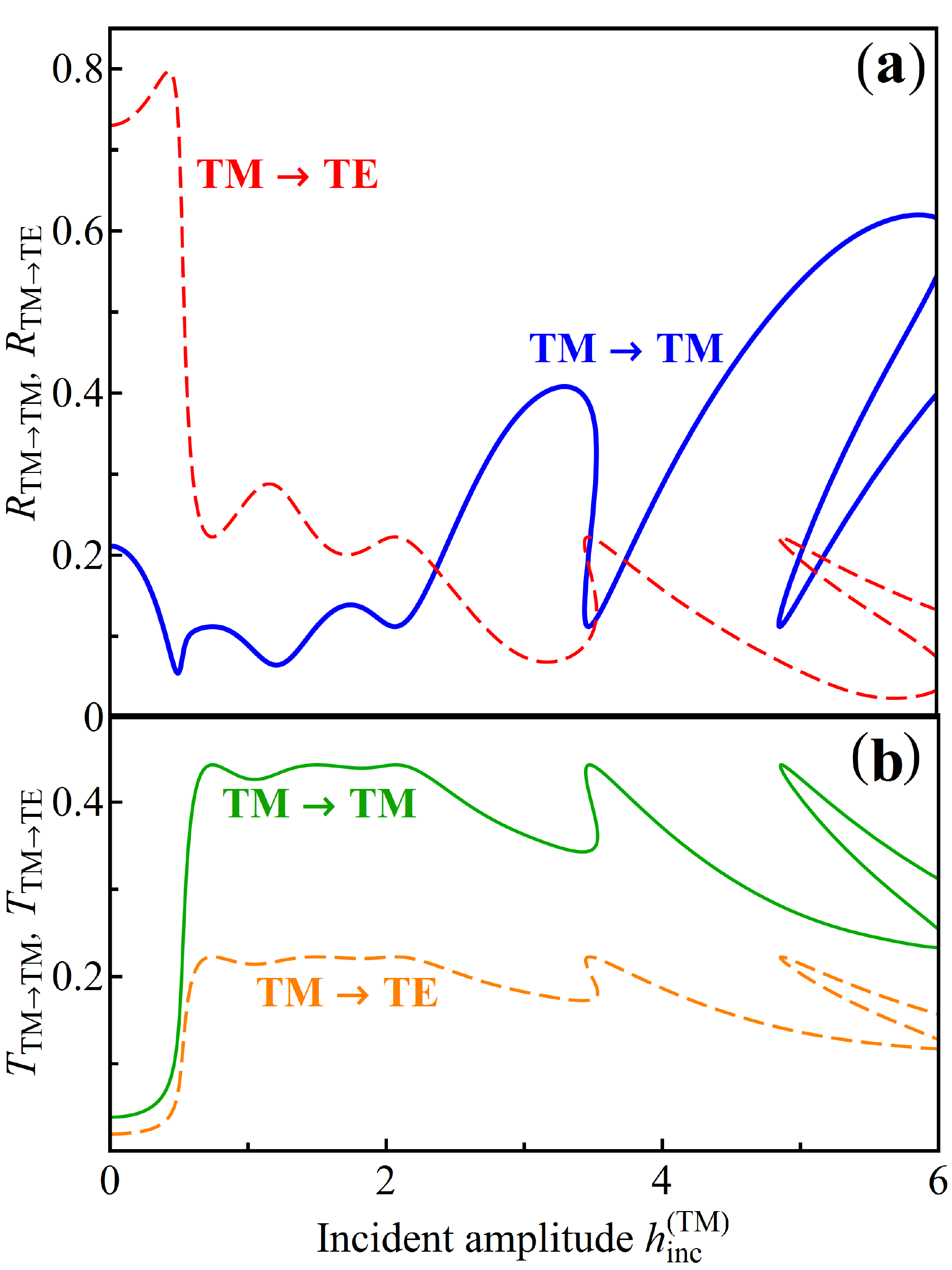}
\caption{(Color online) (a) Reflectance $R_{\rm TM\rightarrow TM}$ (blue solid line) and transformation coefficient $R_{{\rm TM \rightarrow TE}}$ (red dashed line) versus the dimensionless amplitude $h_{\rm inc}^{\rm (TM)}$ of the incident TM wave. (b) Transmittance $T_{\rm TM\rightarrow TM}$ (green solid line) and transformation coefficient $T_{{\rm TM \rightarrow TE}}$ (orange dashed line) versus the dimensionless amplitude $h_{\rm inc}^{\rm (TM)}$ of the incident TM wave. The parameters used here are the same as in Fig.~\ref{colorplots}.}
\label{TM}
\end{center}
\end{figure}

Similarly, we considered the case when the wave incident onto the layered superconductor is TM-polarized, and calculated numerically the reflectance $R_{\rm TM\rightarrow TM}$ and transmittance $T_{\rm TM\rightarrow TM}$ for the TM waves,
\begin{equation}\label{Rtm}
R_{\rm TM\rightarrow TM}=\Big|\dfrac{h_{\rm ref}^{\rm (TM)}}{h_{\rm inc}^{\rm (TM)}}\Big|^{2},
\quad
T_{\rm TM\rightarrow TM}=\Big|\dfrac{h_{\rm tr}^{\rm (TM)}}{h_{\rm inc}^{\rm (TM)}}\Big|^{2},
\end{equation}
as well as the transformation coefficients $R_{{\rm TM \rightarrow TE}}$ and $T_{{\rm TM \rightarrow TE}}$ for the TE waves that appear in the vacuum regions $x<0$ and $x>D$, respectively,
\begin{equation}\label{trans-tm}
R_{\rm TM\rightarrow TE}=\Big|\dfrac{h_{\rm ref}^{\rm (TE)}}{h_{\rm inc}^{\rm (TM)}}\Big|^{2},
\quad
T_{\rm TM\rightarrow TE}=\Big|\dfrac{h_{\rm tr}^{\rm (TE)}}{h_{\rm inc}^{\rm (TM)}}\Big|^{2}.
\end{equation}
Figure~\ref{TM} shows the dependences of these coefficients on the dimensionless amplitude $h_{\rm inc}^{\rm (TM)}$ of the incident TM wave.

\section{Conclusions}

In this paper, we have studied theoretically the reflection and transmission of electromagnetic waves through a finite-length layered superconductor placed inside a waveguide with ideal metal walls. We assume that the superconducting layers are parallel to the waveguide axis. We show that, even in the nonlinear regime, the superposition principle is valid for two waves with mutually-orthogonal polarizations matched to the axis which is perpendicular to both the waveguide axis and the crystallographic \textbf{c}-axis of the superconductor. These two waves do not convert into each other after the reflection from the superconductor, propagate independently, and show principally different behavior. The wave of H-$_\perp$ polarization excites a strong shielding current along the crystallographic \textbf{ab}-plane of the superconductor and, therefore, reflects nearly completely from the superconductor and excites only an evanescent mode inside it. The wave of E-$_\perp$  polarization does not contain the electric field component parallel to both the sample surface and the crystallographic \textbf{ab}-plane. Therefore, it partially reflects and partially transmits through the sample. We have studied nonlinear reflection and transmission of the wave of E-$_\perp$  polarization and shown that the transmittance varies from 0 to 1 when changing the incident wave amplitude. We have also studied the nonlinear interaction and mutual transformation of the transverse electric and transverse magnetic modes in layered superconductors.

\section{Acknowledgements}

We gratefully acknowledge partial support from the RIKEN iTHES project, JSPS-RFBR Contract No.~12-02-92100, Grant-in-Aid for Scientific Research (S), and the Ukrainian-Japanese project ``Josephsonics'' (grant 52/417-2013).


\begin{thebibliography}{99}

\bibitem{Kl-Mu} R.~Kleiner, F.~Steinmeyer, G.~Kunkel, and P.~M\"{u}ller, \prl \textbf{68}, 2394 (1992).

\bibitem{Kl-Mu2} R.~Kleiner and P.~M\"{u}ller, \prb \textbf{49}, 1327 (1994).

\bibitem{Thz-rev} S.~Savel'ev, V.A.~Yampol'skii, A.L.~Rakhmanov, and F.~Nori, Rep. Prog. Phys. \textbf{73}, 026501 (2010).

\bibitem{rev2}X. Hu and S.-Z. Lin, Supercond. Sci. Technol. \textbf{23}, 053001 (2010).

\bibitem{sine-gord} S.~Sakai, P.~Bodin, and N.F.~Pedersen, J. Appl. Phys. {\bf 73}, 2411 (1993).

\bibitem{SG2} S.N.~Artemenko and S.V.~Remizov, JETP Lett. {\bf 66}, 811 (1997).

\bibitem{SG3} S.N.~Artemenko and S.V.~Remizov, Physica C {\bf 362}, 200 (2001).

\bibitem{SG4} Ch.~Helm, J.~Keller, Ch.~Peris, and A.~Sergeev, Physica C {\bf 362}, 43 (2001).

\bibitem{SG5} Ju.H.~Kim and J.~Pokharel, Physica C {\bf 384}, 425 (2003).

\bibitem{nl1} S.~Savel'ev, A.L.~Rakhmanov, V.A.~Yampol'skii, and F.~Nori, Nature Phys. {\bf 2}, 521 (2006).

\bibitem{nl2} V.A.~Yampol'skii, S.~Savel'ev, A.L.~Rakhmanov, and F.~Nori, Phys. Rev. B {\bf 78}, 024511 (2008).

\bibitem{nl3} S.~Savel'ev, V.A.~Yampol'skii, A.L.~Rakhmanov, and F.~Nori, Phys. Rev. B {\bf 75}, 184503 (2007).

\bibitem{nl4} S.S.~Apostolov, Z.A.~Maizelis, M.A.~Sorokina, V.A.~Yampol'skii, and F.~Nori, Phys. Rev. B {\bf 82}, 144521 (2010).

\bibitem{dienst13} A.~Dienst, E.~Casandruc, D.~Fausti, L.~Zhang, M.~Eckstein, M.~Hoffmann, V.~Khanna, N.~Dean,
M.~Gensch, S.~Winnerl, W.~Seidel, S.~Pyon, T.~Takayama, H.~Takagi and A.~Cavalleri, Nature Mat. {\bf 12}, 535 (2013).

\bibitem{Bul_kosh} L.~N.~Bulaevskii, A.~E.~Koshelev, Phys. Rev. Lett
{\bf 99}, 057002 (2007).

\bibitem{Ozyuzer} L.~Ozyuzer, A.~E.~Koshelev, C.~Kurter, N.~Gopalsami,
Q.~Li, M.~Tachiki, K.~Kadowaki, T.~Yamamoto, H.~Minami, H.~Yamaguchi,
T.~Tachiki, K.~E.~Gray, W.-K.~Kwok, U.~Welp, Science,
\textbf{318}, 1291 (2007).

\end{thebibliography}
\end{document}